\documentclass[aps,prfluids,reprint,groupedaddress,showpacs,longbibliography]{revtex4-1}

\usepackage{graphicx}
\usepackage{dcolumn}
\usepackage{bm}
\usepackage{siunitx} 
\usepackage{gensymb}
\usepackage{amsmath}
\usepackage{float}
\usepackage{xcolor}

\begin{document}

\title[Competing gravitational and viscous effects]{Competing gravitational and viscous effects in 3D two-phase flow investigated with a table-top optical scanner}

\author{Joachim Falck Brodin$^1$}
\email[]{jbrodin@online.no}

\author{Per Arne Rikvold$^{1,3}$}
\author{Marcel~Moura$^1$}
\author{Renaud Toussaint$^{1,4}$}
\author{Knut J\o{}rgen M\aa{}l\o{}y$^{1,2}$}
\affiliation{$^1$PoreLab, The NJORD Centre, Department of Physics, University of Oslo, P.O. Box 1048 Blindern, 0316 Oslo, Norway.\\
$^2$PoreLab, Department of Geoscience and petroleum, Norwegian University of Science and Technology, Trondheim, Norway.\\
$^3$Department of Physics, Florida State University, Tallahassee, FL 32306-4350, USA\\
$^4$Universit\'e de Strasbourg, CNRS, ITES UMR 7063, F-67000 Strasbourg, France}

\date{\today}

\begin{abstract}
We present experiments and theory describing the transition from viscosity-stabilized flow to gravitationally unstable fingering for two-phase flow in a 3D synthetic porous medium. Observation is made possible by the use of our newly developed table-top 3D-scanner based on optical index matching and laser-induced fluorescence, which is described in detail. In the experiment, a more dense, more viscous fluid injected at a fixed flow-rate from a point source at the top of the flow cell displaces a less viscous, less dense fluid.  We observe a stable invasion zone near the inlet, which increases in size with increasing flow rates, and presents initially a close to  hemispherical shape. At later times, the invasion front transits to an unstable mode and a fingering flow regime. The transition occurs at a predicted critical radius,  $R_c$, corresponding to the zero of the combined viscous and gravitational pressure gradient. 
\end{abstract}

\maketitle

\section{Introduction}
\begin{figure*}
	\centering
	\includegraphics[width=\linewidth]{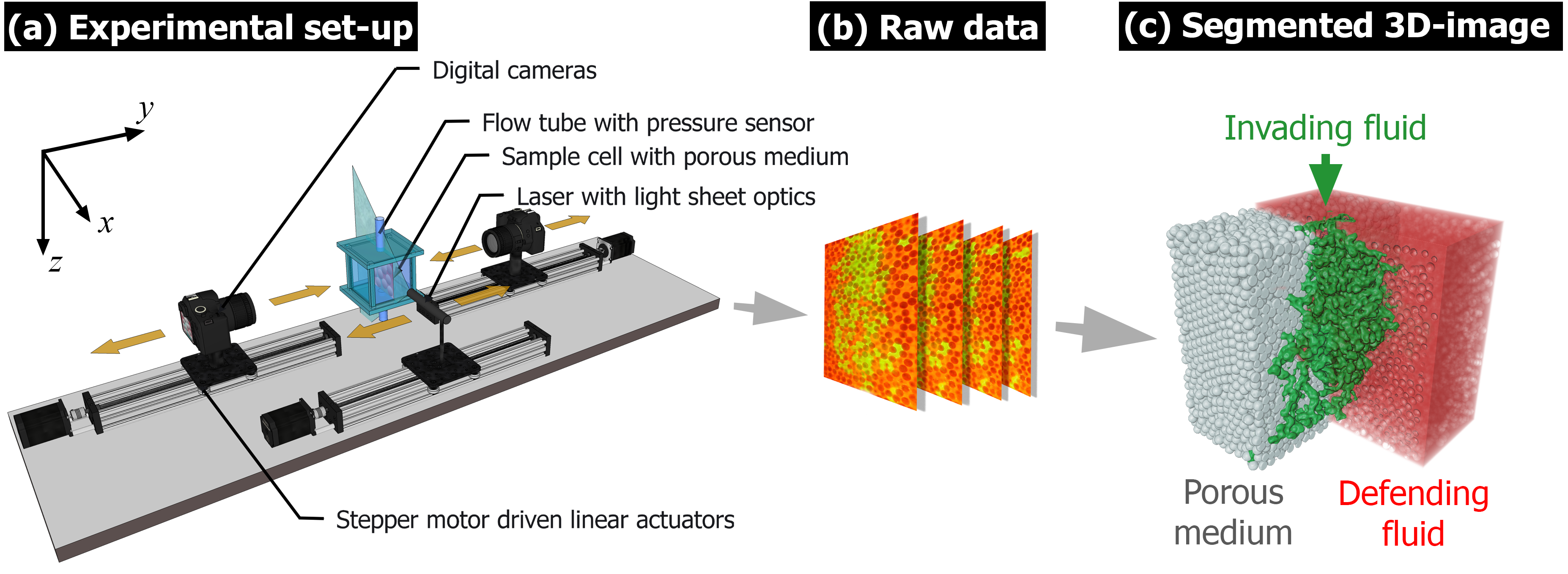}
	\caption{(a) Experimental set-up. (b) Illustration of the unprocessed, raw data. (c) Example of a fully segmented 3D scan from a two-phase flow experiment in an artificial porous medium made from spherical glass beads.}
	\label{fig:diagram_set-up1}
\end{figure*}
Fluid dynamics in complex structures is a rapidly growing research field, considering scales ranging from the micro-scale, such as fluid transport in cells, to the macro-scale, including ocean currents and river networks.  In some cases such systems enable visual study, such as monitoring of river systems through satellite imagery, but often the dynamics are hidden from view, underground or inside opaque bodies. Experimental examination of such processes is and  has been an active field of study for well over a century \citep{darcy1856fontaines}. For simplicity, both experimentally and theoretically, most of the work has been conducted in 2D systems, usually through the adaptation of some form of the Hele-Shaw cell \citep{heleshaw1898}. To study 3D geometries, one can for example use X-rays \citep{Berg2013} or nuclear magnetic resonance \citep{allen1997morphology,yan2012experimental}, but each technique has its limitations, both in terms of what one can image and in how one can interact with the experiment. Variations on other 3D set-ups have existed at least since the 1980s \citep{clement1985multiple,Frette1990}, but access to modern computers, optics, and cameras opens up for methods that previously were practically unfeasible.

In this paper we report results of experiments, in which a viscous, dense liquid is injected from above into a synthetic porous medium, displacing a less dense, less viscous liquid. The competition between viscosity and gravity causes a transition from viscosity-stabilized flow to gravitationally unstable fingering at a critical distance from the injection point. This distance increases with the injected flow rate. Observation of the complex 3D flow patterns is enabled by a newly developed table-top experimental set-up based on optical index matching and laser-induced fluorescence.

The paper is organized as follows. In section \ref{sec:met} we describe the design and configuration of the optical 3D-scanner. In section \ref{sec:exp} we present the first flow experiment conducted with the scanner, including methodology and theory. In section \ref{sec:res} we we present and analyze all results. A discussion and conclusion follow in section \ref{sec:disc}. 

\section{Design and configuration}\label{sec:met}
\subsection{Optical 3D scanner based on index matching}
Our approach employs an underlying principle already in use in several labs \citep{Harshani2017,Moroni2007,datta2014,stohr2003,roth2015,nascimento2019,Ovdat2006,sharma2011,Holzner2011,Heyman,kang2010,kong2011time, Dalbe-Morphodynamics}, based on optical index matching and laser-excited fluorescence, known as RIM (refractive index matching methods) \citep{Wiederseiner2011}. The advantage of our set-up is the hands-on approach, with the experimental dynamics being directly observable by eye, while remaining on a manageable scale, with the whole experiment fitting on a standard-sized table. The set-up can be assembled mostly from available products at a reasonable cost. To put this into context, our set-up costs around 10 000 EUR, which is one to two orders of magnitude less than a CT-system. However, its implementation is rather challenging, with many interconnected components that all must be controlled and accounted for. This article should therefore serve as a reference guide for anyone wishing to construct a similar set-up.

The method relies on images being captured by conventional digital cameras and the use of optical lenses, filters, and laser-induced fluorescence. Although the experiments remain within the domain of flow in porous media, the method relies heavily on optical  principles. As key elements, the set-up uses two cameras to assure the image quality and increase the possible sample size, together with separate actuators for the cameras and the laser to maintain the imaging geometry and allow for constant focus at a wide aperture.

In particular, the method used here differs from methods used in other labs \citep{Harshani2017,Moroni2007,datta2014,stohr2003,roth2015,nascimento2019,Ovdat2006,sharma2011,Holzner2011,Heyman,kang2010,Dalbe-Morphodynamics} in that the cameras and laser are moved by separate linear actuators, allowing for preserved imaging geometry and focus. By also using two cameras we achieve twice the possible scanning depth, allowing for larger systems. Some preliminary experimental results and a theoretical analysis are given in \citep{brodin2020visualization}.

The refractive index of a fluid or solid, $n$, is the ratio of the velocity of light in vacuum to that in the medium. When media are index-matched, they are rendered transparent to one another, and the interfaces between the phases become invisible. This means that light can pass through the medium, unhindered and without distortions. 

Figure \ref{fig:diagram_set-up1}(a) shows a schematic of the experimental set-up. At the center  is a cubic flow cell, containing a synthetic porous medium. It is sealed by glass walls on the four sides, and by plexiglass plates with embedded valves at the center at the top and bottom. At the onset of each experiment, the sample is fully saturated by a fluid with a refractive index that as closely as possible matches that of the porous medium. To the naked eye, the flow cell appears as a homogeneous, solid block of glass, rendering the porous medium invisible. A flow experiment is initiated by injecting another, immiscible fluid with matching refractive index through the valve at the top or at the bottom. To image the flow, we introduce different fluorescent dyes in the fluid phases. The fluorescence is induced by a laser, which illuminates a vertically oriented 2D light sheet. The fluid phases appear as patches with colors corresponding to the dyes. The porous medium remains dark, as it is transparent to the light. The laser is driven in the direction normal to the illuminated sheet, using a stepper-motor driven linear actuator. The illuminated sheet is imaged, frame by frame, by the two digital cameras, which are also moved by linear actuators. The raw data thus constitute a stack of 2D images, as illustrated in Fig. \ref{fig:diagram_set-up1}(b). When organized in sequence, they provide a 3D rendering of the cell, mapping the solid and fluid phases, respectively.

\subsection{Index-matched phases and fluorescent dyes}
\begin{figure}[h!]
	\centering
	\includegraphics[width=\linewidth]{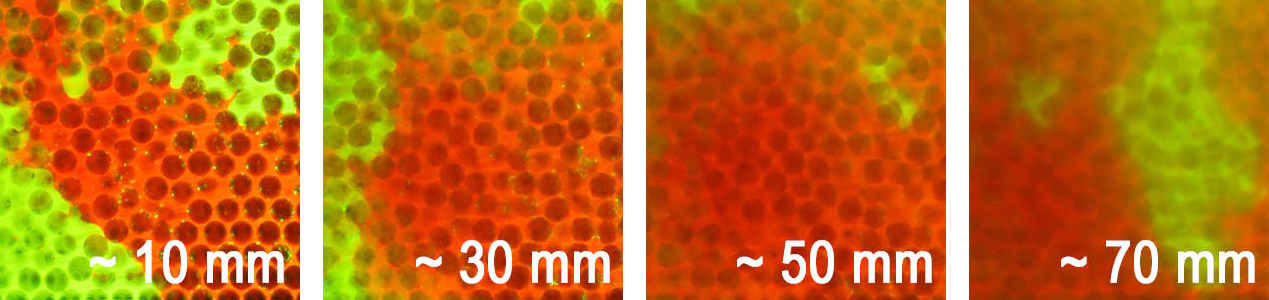}
	\caption{Images taken at different distances into the sample. Here we see two fluid phases, red canola oil and green glycerol, in a medium of randomly packed 3 mm borosilicate glass beads. Because of the imperfect index matching, the image quality deteriorates as the illuminating light sheet is moved further into the sample.}
	\label{fig:depth}
\end{figure}

The experimental approach  depends on the index matching of the solid and the fluid phases. However, in our configuration the refractive indices do not match perfectly, and the cell will contain some contamination. This leads to loss of light intensity and distortions that in turn reduce the distance from which the illuminated sheet can be projected. This sets limits on the image quality and the possible system size. In this particular set-up, these effects are countered by using notch filters, blocking scattered laser light, and two cameras, each of which images one half of the sample. Here it should also be noted that the pore scale  comes into play, as smaller pore size will lead to a higher concentration of interfaces between the phases, which leads to more scattering, and thus to a smaller possible system size.  Figure \ref{fig:depth} demonstrates the effects of imperfect index matching. Close to the sample surface there is little distortion and loss of image quality, but at some distance the image deteriorates significantly.  

Finding an appropriate combination of solid, immiscible fluids, and fluorescent dyes proved to be a  challenge in the development of the experimental set-up. Generally, the refractive index is a function of both temperature and wavelength. The index matching can thus be tuned by controlling the temperature and by choosing a laser and dyes that emit at optimal wavelengths for the selected phases. Some fluids offer the possibility to tune the refractive index chemically, for instance by controlling the fractions in a glycerol-water mixture \citep{Heyman2020}. Tuning is more challenging with multi-phase experiments, as two or more fluid phases will not necessarily match in refractive indices with a solid phase at the same temperature. There are companies that supply tailored fluids for index-matching purposes, but they cater to microfluidics applications and their fluids are too expensive for the table-top scale of the set-up described here. The degree of index matching strongly affects the possible image quality, as well as how far into a sample it is possible to image. It therefore also influences what types of experiments are possible. In the set-up presented here, experimental simplicity is favored, leaving out temperature control and chemical tuning.

The method for deriving a good fluid pair was a mixture of literature search and trial and error (a useful reference is an online library at www.refractiveindex.info). In the end, a determining factor was the ease of handling and the safety of the fluids. For instance, silicone oil was a considered alternative, used in a similar set-up \citep{Dalbe-Morphodynamics}, but it was discarded because of the moderate toxicity and the fact that it makes it very difficult to clean the instruments between experiments. 

To distinguish the porous medium and the fluid phases in the imaging, the fluids are dyed with fluorescent dyes that emit at wavelengths as widely spaced as possible, thus appearing as patches of different color.  The fluids and dyes used in the experiment described here, are glycerol dyed with 10-50 mg/l Fluorescein emitting green light at 548 nm, and vegetable oil dyed with 1-10 mg/l Pyrromethene emitting red light at 650 nm, both dyes supplied by Exciton. The fluid properties are summarized in table \ref{table:fluidMeas}, in section \ref{sec:exp}. To illuminate the cell, we use a 532 nm ZM18GF029 laser from Z-laser, pre-fitted with a prism that spreads the beam out in a 60$^{\circ}$ sheet of thickness 0.3 mm. The porous medium remains transparent to the exiting light, appearing as dark patches.

\subsection{Imaging and data acquisition}
The scanner connects to a computer with 128 GB of RAM and a very fast SSD disk. The mechanical operation of the linear actuators and imaging is handled by a single script. In this configuration the code is written in Python.  The image data are captured through straight-forward optical imaging with standard optical lenses and cameras. In the configuration described above, there are two MER-230-168U3C  2.3MP RGB industrial vision cameras with USB-3.0 interfaces, from Get-Cameras, with 50 mm f/2.4 lenses from the same provider. The lenses are fitted with NF533-17  notch filters from Thorlabs, which filter out scattered light and reflections at wavelengths around that of the laser.  These cameras have the advantage that they can be controlled directly from a script that also handles data acquisition and storage.

A full 3D scan accumulates a large amount of data. Each 3D image consists of $1000\times1000\times1000$ voxels in 8-bit 3-channel color (RGB). The raw data for each image requires about 7 GB of space that has to be either cumulatively buffered in RAM, or written to permanent storage between each scan. Therefore, the image resolution determines the possible frame rate of the scanner. We opted to configure the scanner so that the voxel resolution would be identical in all three dimensions, making a combination of the possible frame rates of the cameras and the pixel dimensions on the sensors determines the final integration time of the full scan. The cameras used can operate at up to 168 frames per second (fps), but buffering to the computer, generating a live video feed, and minimal processing limit the practical frame rate to 50 fps. A full image of 1000 sequential 2D images takes 20 seconds to record. In addition, writing the image to disk and resetting the laser and camera positions bring the maximum scan rate to about one full image per minute. It should be noted that this rate can be increased by reducing the image resolution or buffering all data to RAM, and by recording to a hard-disk after the experiment. Because of the high viscosity of glycerol, see Table \ref{table:fluidMeas}, leading to relatively slow dynamics, it was deemed that the scan rate was sufficiently high at 1 scan per minute. It should also be noted that the 3D images are very time consuming to process and analyze, so that gathering larger amounts of data would lead to extended processing time and require corresponding available storage space.

The imaging geometry is determined by illuminating the center cross section of the sample and shifting and refocusing one of the cameras until a desirable cropping of the sample is achieved. The other camera is then moved to an equal distance to the other side of the illuminated sheet. By measuring the pixel dimensions of an object of known size, the actual dimension of an imaged pixel can be determined. 

The cameras follow the motion of the laser, but to compensate for the magnification of the illuminated plane, caused by the image being projected through a medium of higher refractive index, $n$, than the air between the camera lens and the sample, the laser moves at a constant speed equal to $n$ times the speed of the cameras. In this manner a constant optical path is preserved, and the imaging magnification is kept constant. The velocities for the laser and cameras are set so that two successive 2D images correspond to focus on planes separated by one pixel dimension, i.e. they are given in Eq. (\ref{eq:vel}).

\begin{equation}
\begin{aligned}
v_{laser}&=\textit{pixel dimension}\times \textit{camera fps}\\
v_{cameras}&=v_{laser}/n,
\end{aligned}
\label{eq:vel}
\end{equation}

\noindent where $pixel$ $dimension$ signifies the physical size of the individual pixels on the camera sensor. This ratio of camera speed to laser speed is an essential element of the set-up, as the constant optical path keeps the scaling of the imaging and allows the use of a wide aperture, which in turn leads to a shallow imaged depth of field, focused solely on the plane illuminated by the light sheet.

At the onset of each scan, the laser, on its carriage on the linear actuator, is outside of the sample, close to one of the vertical glass faces. The two cameras are in their corresponding positions. A scan is conducted by driving the laser sheet through the sample. The camera nearest the side the laser is in starts recording to the computer RAM, filling out a pre-initiated array. When the light sheet reaches the center of the sample, the second camera takes over. As the sheet reaches the end of the cell, the data array is written to the HD while the carriages are sent back to their original positions. 

\subsection{Image processing and segmentation}
The scanner images conveys an intuitive mapping of the phases to our eyes, but translating this into an automated segmentation algorithm remains challenging. As can be seen in Fig. \ref{fig:depth} in section \ref{sec:disc} such an algorithm has to deal with gradients and fluctuations in light intensity and image quality. There is no unique way to do this, but rather a whole range of possible avenues, ranging from standard image processing, to using machine learning or neural networks.

\begin{figure}[H]
	\centering
	\includegraphics[width=1\linewidth]{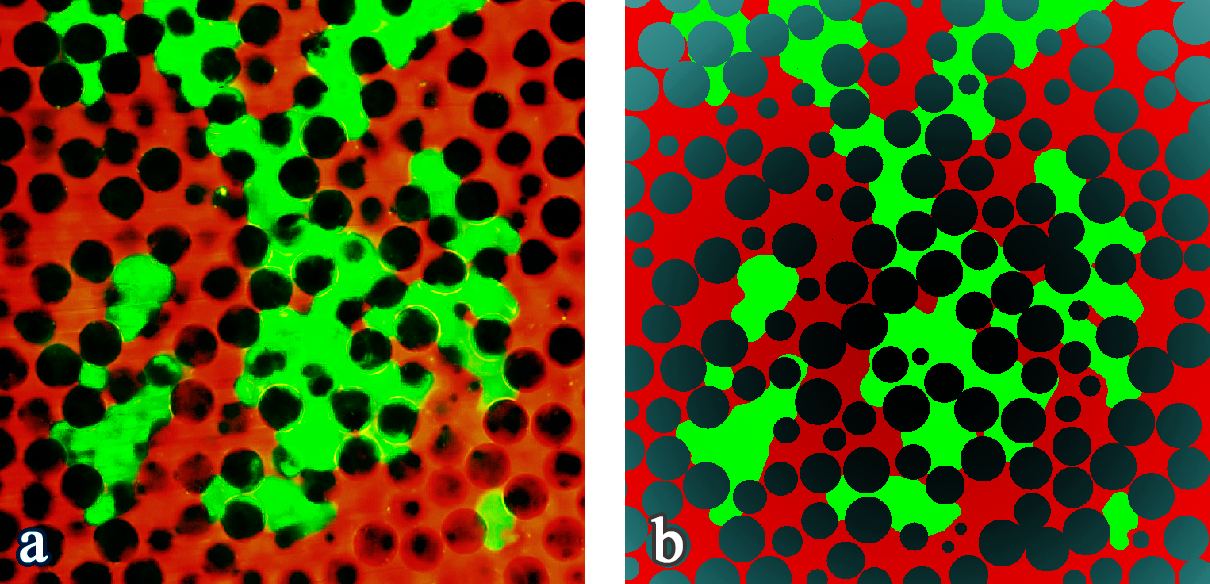}
	\caption{(a) Raw data. (b) Segmented image.}
	\label{fig:segmented}
\end{figure}

For this paper, we have used a combination of image processing with the OpenCV Python library and segmentation, analysis, and visualization with the Amira Avizo software from Thermo Fisher. The raw data consists of a 3D matrix of voxels that each have separate 8-bit RGB values. There are gradients in light intensity and image quality, as well as noise and artifacts, that have to be considered and adjusted for.  The steps taken are as follows.
\begin{itemize}
\item \textbf{Image alignment:} The two cameras record the sample from opposite sides, so the recorded data from the second half needs to be flipped to match that from the first half. Although the cameras are carefully aligned, there is a possibility of slight misalignment. A transformation function, based on the findTransformECC function from OpenCV, is made by comparing the center-most images from the two cameras. This function is then applied, through the warpAffine function from OpenCV, to all the images recorded by the second camera, to rotate and stretch the image, so that the geometry is preserved throughout the sample.
\item \textbf{Segmentation:} The data are loaded into Avizo. The segmentation contains two main steps. First, we segment out the solid phase, using a scan taken prior to the flow experiment, with the cell fully saturated with the defending phase. As the fluids contain red and green dyes respectively, this analysis is carried out solely in the image channel for the color corresponding to the defending fluid.  We localize the center of mass of each glass bead and generate solid spheres with the appropriate diameter at the locations. This image of the solid matrix is then used in the segmentation of the ensuing scans. Second, the invading fluid is segmented out from the color channel corresponding to the invading fluid dye. The defending fluid is then the remaining volume.  The result is a full digital mapping of the solid and the two fluid phases. Figure \ref{fig:segmented} shows an example of how the same image looks before and after segmentation.
\end{itemize}

\section{Experiment - competing viscous, capillary, and gravitational forces }\label{sec:exp}
The first experiment conducted with the set-up was a study of the interplay of viscous, capillary and gravitational forces, see Fig. \ref{fig:geometry}. A cubic flow cell, measuring 8 cm $\times$ 8 cm $\times$ 8 cm, was initiated with a fixed medium made up of quasi-uniform borosilicate glass spheres of 3 mm diameter supplied by Sigma Aldrich, with a reference refractive index of $n\approx 1.47$, according to the supplier's data sheet. The cell was sealed on all sides, with an inlet valve, protruding 15 mm into the matrix on top, and an outlet valve at the bottom.  At the onset, the cell was fully saturated by the oil, with a red dye. We injected pure glycerol, with a green dye,  at fixed flow rates from the top. The outlet was connected to a large basin, providing a quasi-fixed pressure at the outlet.

We conducted experiments at different flow rates, and analyzed the obtained invasion patterns. The properties of the fluid phases are summarized in Table \ref{table:fluidMeas}.

 \begin{table}[H]
\caption{Reference values and experimental data for the two fluids, centered at approximately 20$^{\circ}$ C, with the span in viscosity representing $\pm 2^{\circ}$ C.}
\centering
\begin{tabular}{lll}
\toprule
Reference data         &Glycerol         &Rapeseed Oil         \\
\hline
$n$ &    1.46-1.48   &    1.47    \\
\hline
Experimental data&&\\
\hline
$\rho$       &1.26 $\pm$ 0.02 g/cm$^3$                   &0.91 $\pm$ 0.02 g/cm$^3$ \\
$\mu$ &   1350 $\pm$ 140 mPa$\cdot$s                    & 71 $\pm$ 7 mPa$\cdot$s \\
$\gamma$ (vs air)     & 59.9  $\pm$ 1.0 mN/m                & 31.4  $\pm$ 1.0 mN/m \\
$\gamma$ (G vs RO)     & 16.4 $\pm$ 1.0 mN/m               & 16.4 $\pm$ 1.0  mN/m\\
\hline
\end{tabular}
\label{table:fluidMeas}
\end{table}

\subsection{Theoretical context}

As a direct parallel to experiments considered in 2D systems \citep{birovljev1991gravity,Frette1992a,Lovoll2005,toussaint2012two}, our theoretical starting point is that the ensuing dynamics are mainly governed by an interplay between forces dictated by the fluid properties, the wetting conditions, the geometric configuration of the matrix, gravity, and the invasion method. Depending on these parameters, the invading fluid displaces the defending fluid on a scale ranging from full displacement leading to near total saturation, to unstable branched displacement leading to low saturation. The invasion body assumes a geometrical shape which retains characteristics that can be quantified by saturation and topological features. This geometrical structure is the main result delivered by the scanner. We distinguish between the high saturation scenario, with an invasion front preserving the symmetries imposed by the injection geometry and gravity, which we call \textit{stable} displacement, or low saturation, spontaneous symmetry breaking of the front and fingering, \textit{unstable} displacement. For instance, if there is a density contrast between the fluids, the more dense fluid front will be gravity stabilized when invading upwards, and destabilized when invading downwards. A wider spread in pore-throat size distribution will lead to larger fluctuations in the front, as the flow will favor  the wide passages and leave behind pores accessible only through narrow pore throats.

In two-phase flow in a porous medium, the condition that governs if a given pore is invaded or not is whether the pressure difference between the two sides of the interface exceeds the capillary pressure threshold set up by the surface tension, $\gamma$, between the fluid phases in the local pore geometry.  

For single-phase flow in a porous medium  Darcy's law on a differential form, may be written as

\begin{equation}
\vec{v}=-\frac{\kappa}{\mu}(\nabla p- \rho \vec{g}),
\label{eq:DarcyDiff}
\end{equation}
where $\vec{v}$ is the filtration velocity, $\rho$ is the fluid density, $\vec{g}$ is the gravitational acceleration, $\mu$ is the fluid viscosity, $\nabla p$ is the pressure gradient, and $\kappa$ is the permeability of the medium.

For a single fluid phase entering with a constant flow rate, $Q$, at depth $z=0$ into a vertical pipe containing a porous medium with cross section $A$, Darcy's can be rewritten on a integral form as

\begin{equation}
p(z) =p(0)+\left(\rho g-\frac{\mu Q}{\kappa A}\right)z,
\label{eq:Darcy1}
\end{equation}  
where $p(z)$ is the pressure at depth  $z$. Here $\vec{z}$ is pointing downwards, parallel with $\vec{g}$. This gives a local pressure, at depth $z$, from gravity, as well as a viscous contribution.  

\begin{figure}[h!]
	\centering
	\includegraphics[width=0.8\linewidth]{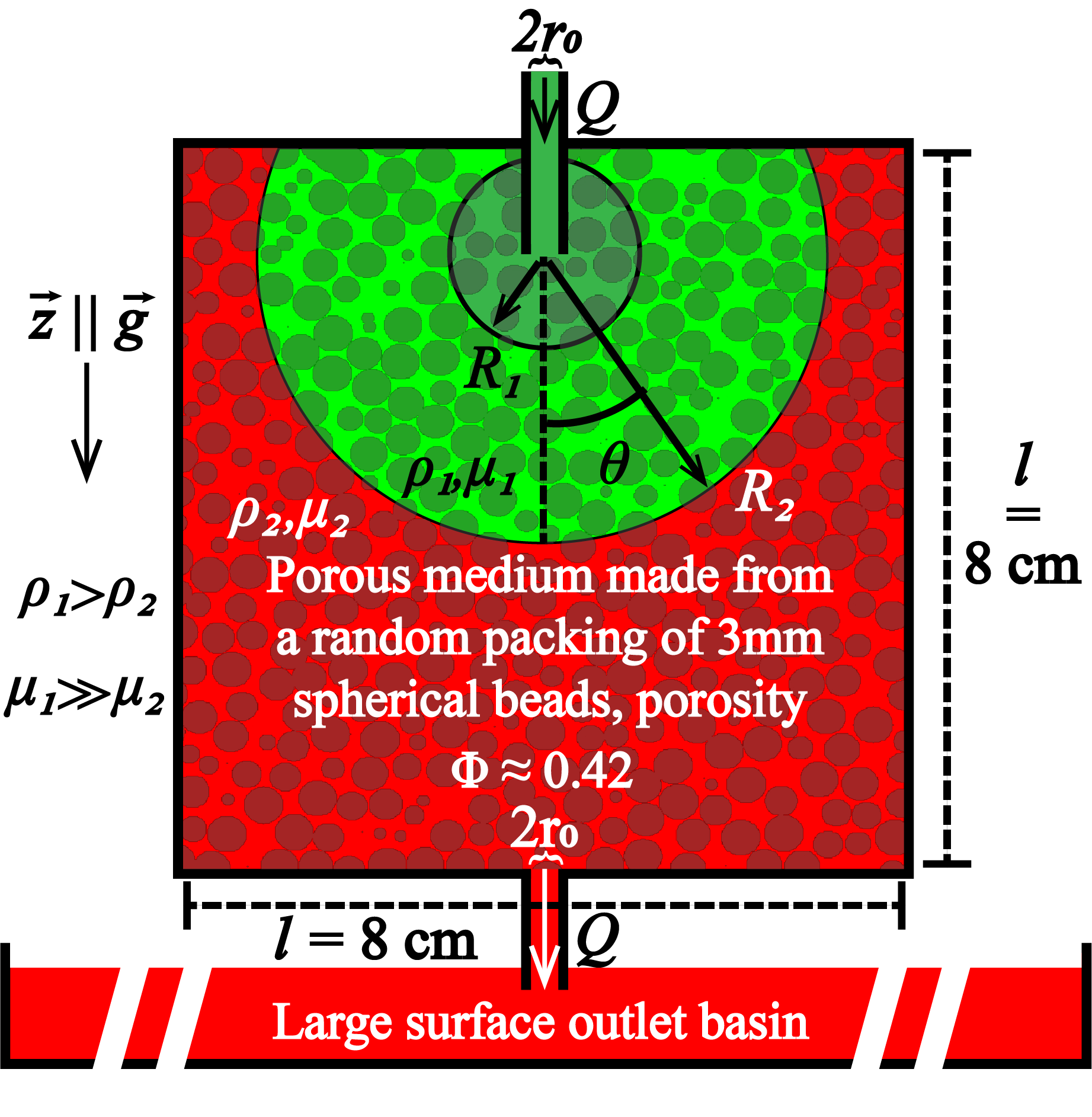}
	\caption{Simplified diagram of the experiment. The cell is fully saturated by liquid 2 (red) at the onset and invaded by liquid 1 (green) from the top, with a constant flow rate, $Q$. The outlet leads to a large-surface reservoir open to the atmosphere, in order to maintain a near constant pressure at the outlet. According to our approximate theory, the invasion starts as a sphere, $R_1$, until the front hits the roof, where the front gradually changes to a downward-pointing hemisphere, $R_2$.}
	\label{fig:geometry}
\end{figure}

In this experiment, invasion occurs from above with a more viscous, more dense fluid. In this situation, gravity is expected to destabilize the invasion, in competition with the stabilizing viscous pressure drop \cite{sabet2021scalings}. In the case of high local fluid velocity, we expect stabilization of the front \citep{birovljev1991gravity,Frette1992a,Lovoll2005}. Due to mass conservation, the fluid velocity decays as a function of distance to the inlet. Based on this, we derive a theory in which we assume a stable invasion near the narrow inlet, caused by the high local velocity. The point injection with gravitational field leads to a  cylindrical symmetry of the problem around a vertical axis aligned with the injection point, far enough from the side boundaries. Close to the inlet where velocity is high, viscous pressure gradients  dominate over gravitational pressure gradients, and the velocity flow is expected to be symmetric under rotation around the injection point. Indeed, from experimental observations, we find it reasonable to approximate the shape of the invading fluid during this stable phase as a downwards-facing hemisphere centered on the inlet. See Fig. \ref{fig:geometry}.

We note that the wetting condition is a natural feature to consider in this context. We conducted measurements of fluid-fluid interface tension and fluid-solid wetting angles with a drop shape analyzer and found the wetting conditions to be indeterminably neutral. We observe that the invading glycerol appears to be the non-wetting phase when it is the invading phase, going through the pore-throats with a convex front, whereas the oil also assumes a convex front shape, when it is used as the invading phase. The wetting condition have therefore not been included in our theory.

\subsection{Permeability Measurements}
\begin{figure}[h!]
	\centering
	\includegraphics[width=0.5\linewidth]{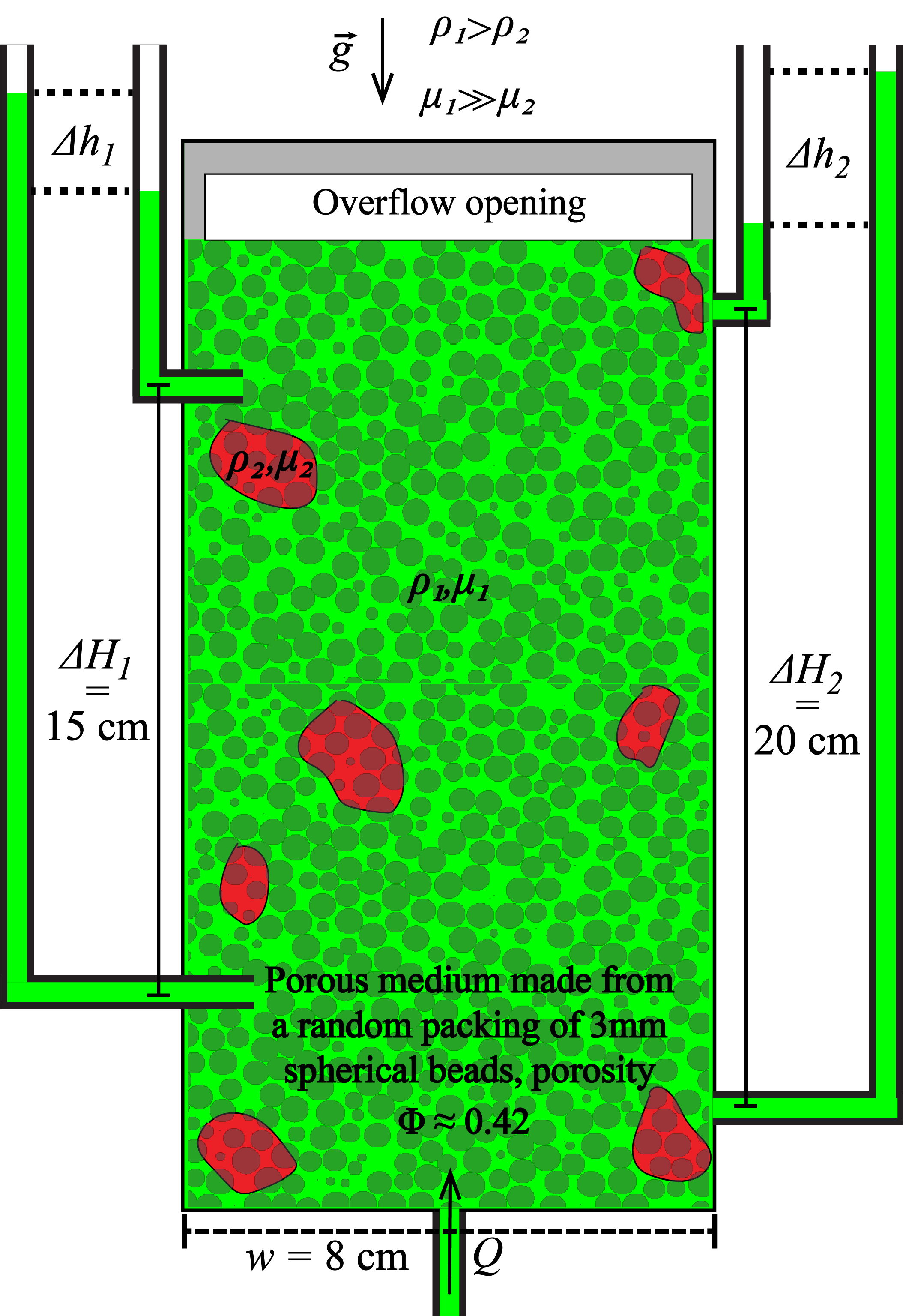}
	\caption{Set-up for permeability measurements. The cell is open at the top, with an overflow to maintain near-constant pressure. Fluid is injected through an inlet at the bottom, with constant flow rate maintained a syringe pump. Holes with nipples are made over two fixed intervals for preassure measurements, one with the nipple at the cell wall, and one with the nipple inserted a couple of centimeters into the matrix, to verify that the pressure gradient is the same inside the matrix, as along the walls. Pressure differences are read off over the intervals as hydrostatic fluid columns in transparent tubes, attached vertically over a ruler to measure height differences, $\Delta h$, as seen on the right. The viscous pressure gradients then can be calculated as $|\nabla p|=\rho_1 g \Delta h/ \Delta H$.  }
	\label{fig:permCell}
\end{figure}

The permeability in the experimental context is a complex issue. Neither the traditional one-phase permeability, as defined by Darcy in Eq. (\ref{eq:Darcy1}) nor relative multi-phase permeability as it is traditionally defined \citep{johnson1959calculation} applies fully to this case, as the fluid saturations evolve and the fluid interface configurations are not in a steady state. To get a best approximation we have made a set-up, quite along the lines of Darcy's experiment, as illustrated and explained in Fig. (\ref{fig:permCell}).

Two sets of measurements were made, one for single-phase flow, and one for two-phase flow. For the single-phase flow, the cell was pre-saturated with glycerol, and the same fluid was injected at a fixed flow rate, $Q$ at the bottom. For the two-phase flow, the cell was pre-saturated with the oil, after which glycerol was injected until where the read-off tubes were all in direct contact with the invading glycerol. Any remaining oil in the tubes was removed with a syringe. Two-phase measurements were then made with glycerol injected at a fixed flow rate at the bottom. The permeability was calculated from Eq. (\ref{eq:Darcy1}) as

\begin{equation}
\kappa=\frac{v\mu_1}{|\nabla p|}=\frac{Q\mu_1\Delta H}{A \rho_1 g \Delta h}.
\end{equation}
The results are presented in Table (\ref{table:perm}).

 \begin{table}[H]
\caption{Results from single-phase and two-phase permeability measurements. The error is evaluated as the standard deviation in the measurements.}
\centering
\begin{tabular}{|c|c|}
\hline
Single-Phase  Permeability       &Two-Phase Permeability\\   
\hline
$\kappa_{SP} =(17\pm3)\times 10^{-9}$ m$^2$ &    $\kappa_{TP}=(8.6\pm0.2)\times 10^{-9}$ m$^2$\\
\hline
\end{tabular}
\label{table:perm}
\end{table}
\noindent As a reference, the Kozeny-Carman relation \citep{Dalbe-Morphodynamics} puts the permeability at $\kappa\approx\frac{r^2}{45}\frac{\phi^3}{(1-\phi)^2}=1.1\times10^{-8}$ m$^2$, where $r=1.5$ mm is the bead radius, and $\phi\approx0.42$ is the porosity measured from the experiments. 

\subsection{The critical radius}
In a real porous medium the front will never be  a perfect hemisphere, but will have fluctuations due to the variations in pore sizes with corresponding capillary pressure-threshold fluctuations. We therefore consider a small perturbation of the hemisphere, corresponding to an interface in a pore located  a distance $a$  in front of the hemisphere.  We consider the two points, $z=R$ on the hemisphere and a perturbation at $z=R+a$ just in front of that point, representing a pore that has just been invaded.  We compare the capillary pressure in this pore, with that at a point on the hemisphere at $z=R$. As an approximation using Eq. (\ref{eq:DarcyDiff}) in polar coordinates for the hemisphere,  the change of pressure in fluid 1 and fluid 2, respectively, can be expressed as
\begin{equation}
\Delta p_1=p_1(z=R+a)-p_1(z=R)=\left(\rho_1 g-\frac{Q \mu_1}{A(R) \kappa_{TP}}\right)a \; ,
\end{equation}
\begin{equation}
\Delta p_2=p_2(z=R+a)-p_2(z=R)=\left(\rho_2 g-\frac{Q \mu_2}{A(R) \kappa_{SP}}\right)a \; ,
\end{equation}
where $A(R)=2 \pi R^2$ is the area of the hemisphere. Let $p^c=p_1-p_2$ be the capillary pressure across the interface between fluid 1 and fluid 2 at a given position. The increase in capillary pressure is then
\begin{equation}
\begin{aligned}
\Delta p^c&=p^c(z=R+a)-p^c(z=R)\\
&=\Delta p_1- \Delta p_2\\
&=\left[(\rho_1-\rho_2) g-\left(\frac{Q \mu_1}{A(R) \kappa_{TP}}-\frac{Q \mu_2}{A(R) \kappa_{SP}}\right)\right]a
 \; .
\end{aligned}
\end{equation}
The capillary pressure at the perturbation  will have an increase due to the hydrostatic pressure and a decrease  due the viscous pressure  corresponding to the first and the second term in the equation above, respectively. If $\Delta p^c>0$, the gravitational term dominates and the invasion is more likely to continue at the perturbation, than at the lagging front. The front is thus rendered unstable. Conversely, if  $\Delta p^c<0$, the viscous term dominates and the front remains stable. We therefore use $\Delta p^c=0$, with the front area approximated as $A\approx 2 \pi R_c^2$, to predict the characteristic length scale $R_c$ for the transition between the viscous stable and the gravitationally unstable displacement. 
\begin{equation}
     R_c=\sqrt{\left[\frac{\mu_1}{\kappa_{TP}} - \frac{\mu_2}{\kappa_{SP}}\right]\frac{Q}{2\pi \Delta\rho g}}.
\label{eq:Rc}
\end{equation}
We note that since $\mu_1 \gg \mu_2$, the $\frac{\mu_2}{\kappa_{SP}}$ term barely contributes to the estimated values for $R_c$.

The theory described above predicts a length scale for the transition from stable, viscosity-dominated invasion to unstable, gravity-driven invasion. However, it does not include the role of the spatial fluctuations in the porous medium. We have described these effects in greater detail in \citep{brodin2020visualization}. As the invasion front progresses, the porous medium presents a distribution of capillary pressure thresholds due to the pinned menisci in the pore throats.  As long as $-\Delta p^c$ is much larger than the characteristic width of the capillary pressure threshold distribution $W_t$, nearly every pore will be  invaded. The  width of stable invasion fronts in quasi 2D experiments has been found to  depend  on the generalised fluctuation number $F=-\Delta p^c/W_t$ with a transition to unstable fronts  when $F=0$ \cite{wilkinson1984percolation, meheust2002,birovljev1991gravity,toussaint2005influence, Maloy2021}.  In our experiments $F=0$  corresponds to the situation where the invasion structure  reaches the  critical radius  where the  local fluctuations  are felt, and the wider pores are preferred over narrower ones.   This leads to fingering  and unstable displacement, such as investigated in 2D \cite{vasseur2013flow}.

\section{Results}\label{sec:res}

\begin{figure}[H]
	\centering
	\includegraphics[width=1\linewidth]{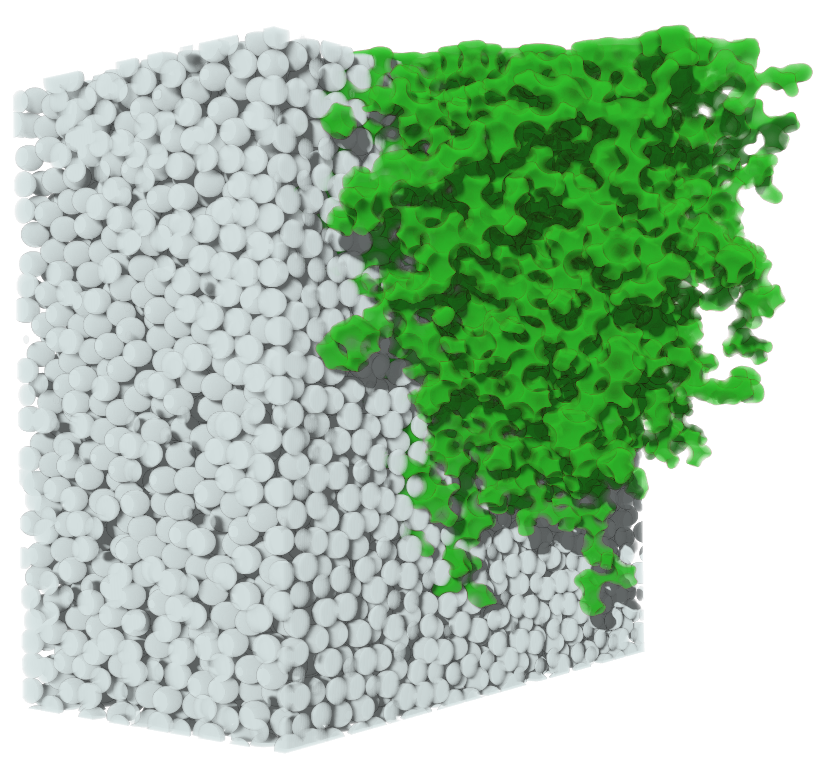}
	\caption{3D image from the scanner, from an experiment with $Q=0.8$ ml/min, captured at the moment the invading fluid reached the bottom of the cell, after 31 minutes. The porous medium is cut in half, with the right side removed for visual access to the invasion body. }
	\label{fig:beadsGly}
\end{figure}

Figure \ref{fig:beadsGly} shows a 3D rendering from a scanner image. The scanner images allow for analyses from the pore scale up to the meso scale of the full flow cell, with the limits set by fine tuning of the refractive indexes and a bigger system. 
 
\subsection{Flow Experiments}
\begin{figure*}
	\centering
	\includegraphics[width=1\linewidth]{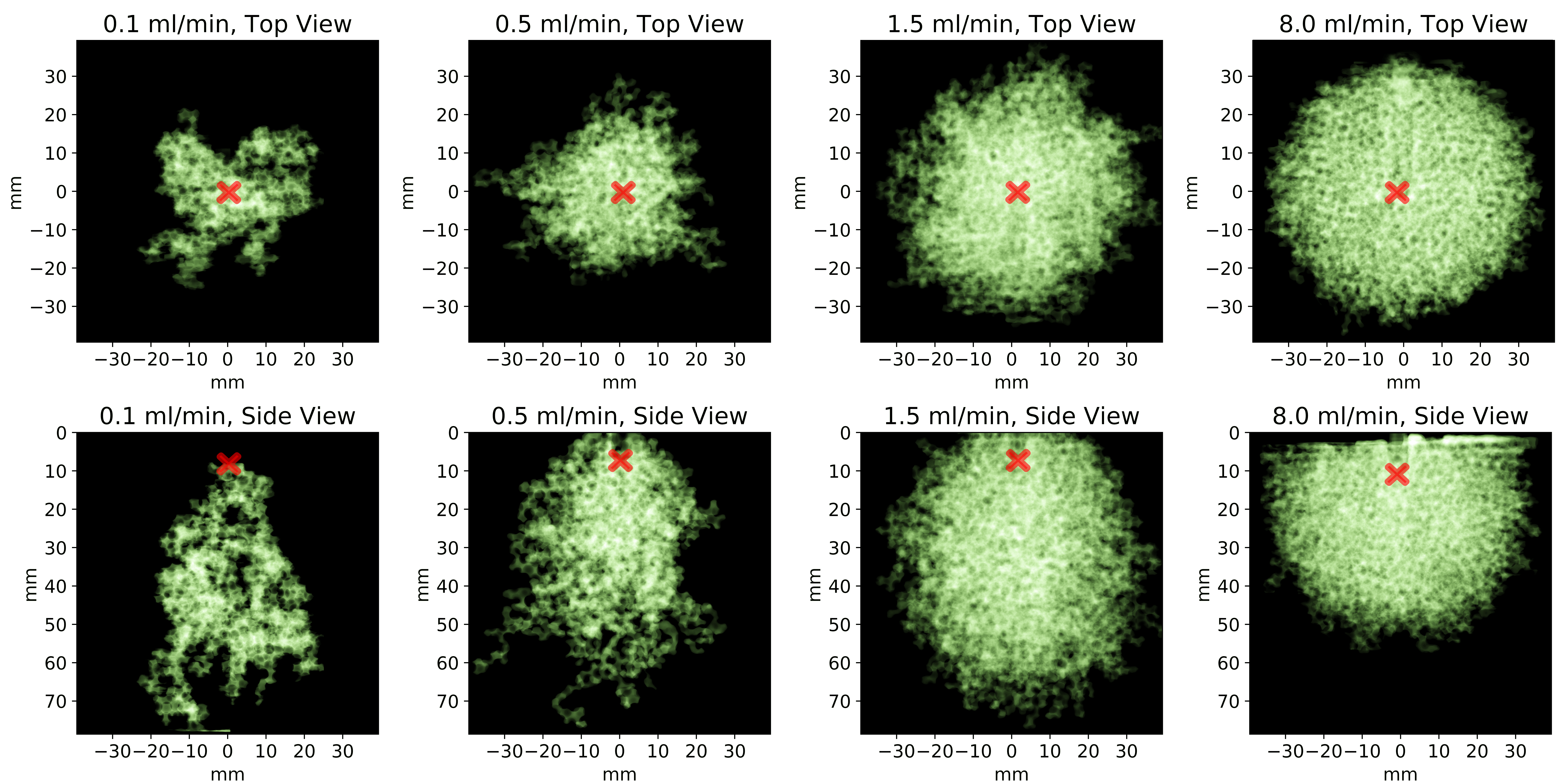}
	\caption{Density projections from the scanner images of the invasion body for four different flow rates. All the images are taken at the moment the fluid reaches either the bottom or the wall of the flow cell. The top row is seen from above and the bottom row is seen from the side. The projections are made by adding the layers of the 3D image, mapping each layer with ones for the invading fluid, and zeros elsewhere, into the image plane.  As can be seen, the body becomes denser, with higher global saturation, as the flow rate is increased. The red crosses indicate the location of the inlet.}
	\label{fig:projections}
\end{figure*}

We conducted fourteen runs, in which we increased the flow rate from 0.1 ml/min to 8.0 ml/min. Each experiment was conducted after resetting the flow cell, with a new bead configuration for the porous medium. Based on our assumptions, the stabilizing viscous pressure drops radially  from the inlet in the invading fluid. This should lead to the flow being unstable, dominated by gravitational forces beyond the critical radius, $R_c$ \citep{auradou1999competition,Meheust2002a,yan2012experimental,vasseur2013flow,Maloy2021}. Figure \ref{fig:projections} shows density projections of the 3D images from four of the experiments. As can be seen, the structures become denser and appear to assume a hemispherical shape as the flow rate is increased.  

\begin{figure}[h!]
	\centering
	\includegraphics[width=1\linewidth]{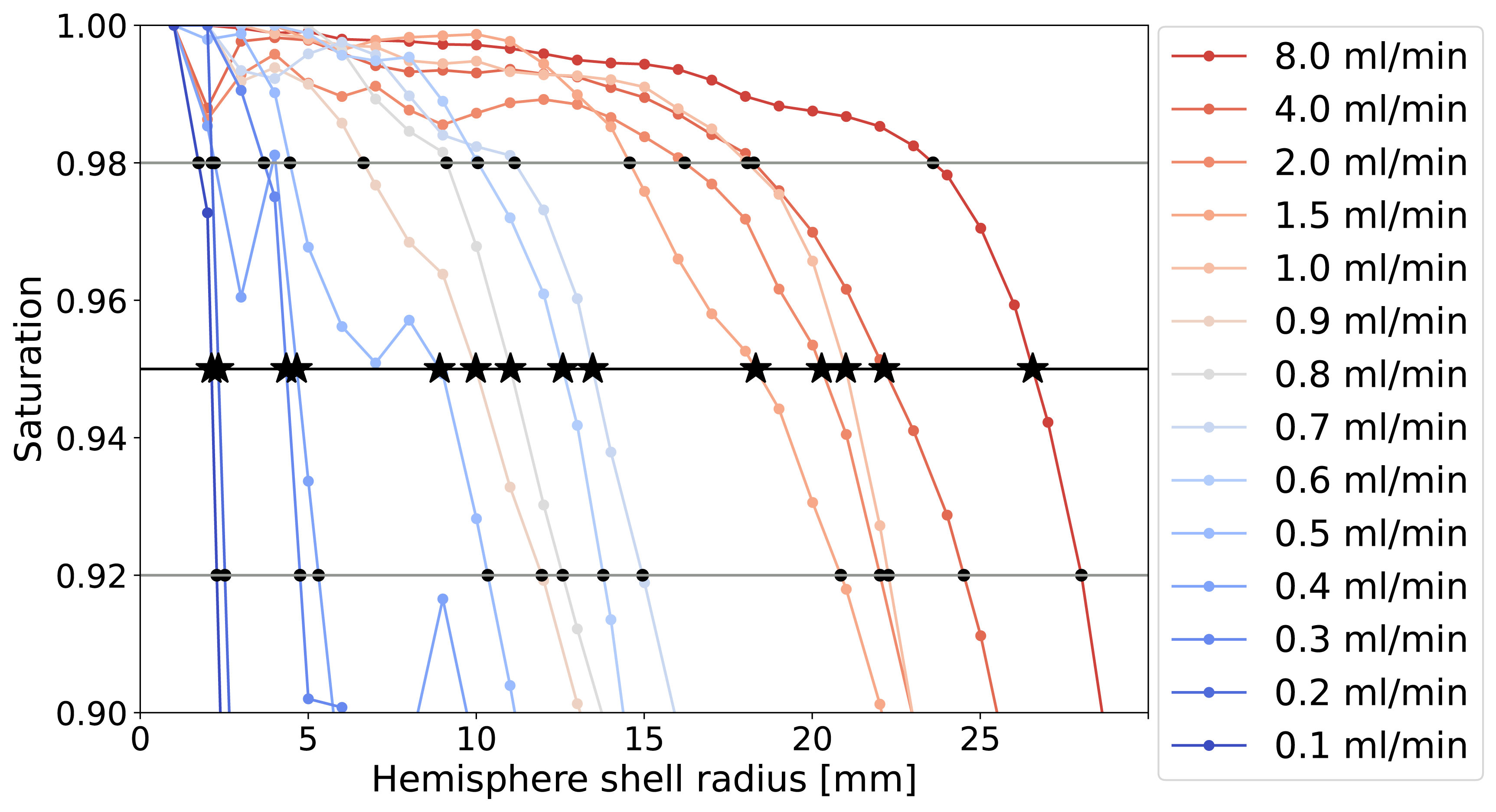}
	\caption{Plots of the saturation inside a downward-facing hemisphere with center at the inlet, imposed over the invasion structures. The points indicate the saturation of the invasion body inside 1-mm incremental hemispherical shells. The black stars are the measured transition radius,$R_c$,  corresponding to a set threshold of saturation, $S=0.95$. The black dots are values corresponding to $S=0.92$ and $S=0.92$, which we use as the error range for  $R_c$. }
	\label{fig:radiusPlot}
\end{figure}

To identify a critical radius, $R_c$, as in Eq. (\ref{eq:Rc}), we consider the saturation near the inlet. In Fig. \ref{fig:radiusPlot} we show a plot of the saturation inside downwards facing hemispherical shells, centered at the inlet. We measure $R_c$ as the radius where the saturation in the shell drops below 0.95. These plots support the assumption that the viscous stabilization only becomes significant beyond a given flow rate.

\begin{figure*}
	\centering
	\includegraphics[width=1\linewidth]{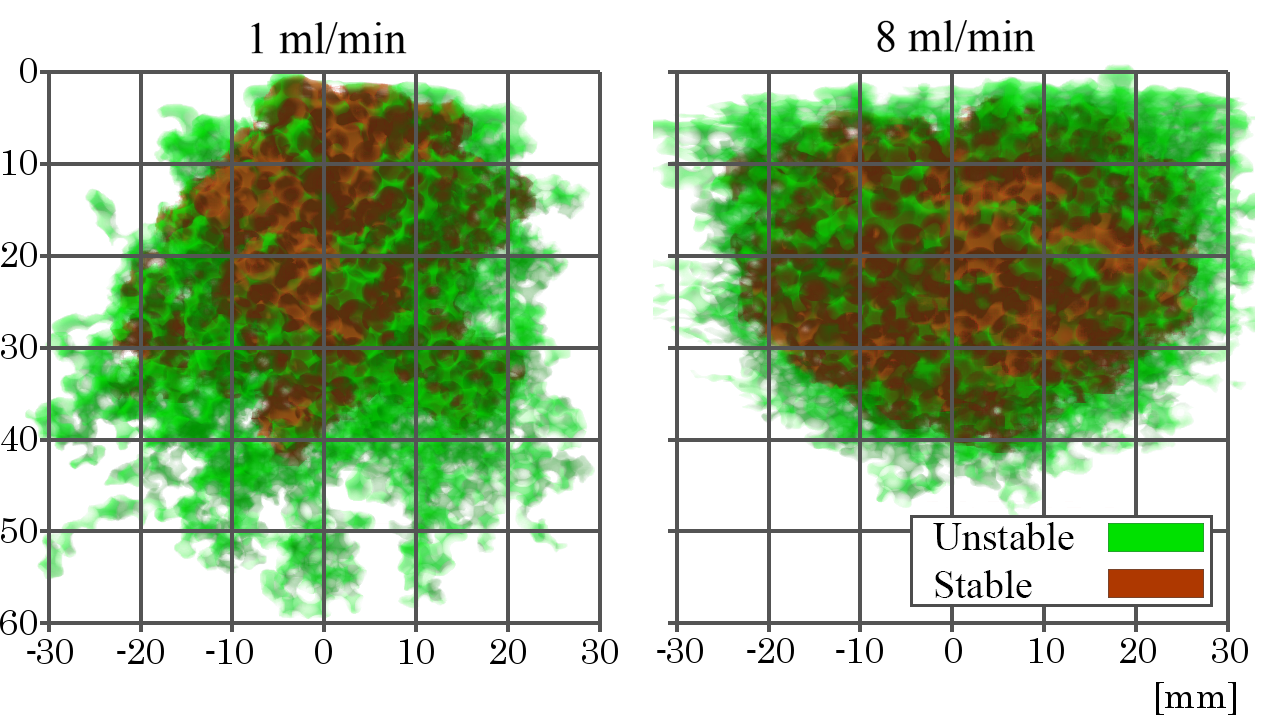}
	\caption{3D renderings of the scanner images of the invasion body for two different flow rates with approximately 23 ml injected in both cases. The green signifies the outline of the full body, and the red inside is segmented out as the \textit{stable} part of the same body. The segmentation procedure, known as morphological opening (erosion-dilation), is a common image-processing procedure, for instance described in \citep{said2016study}. }
	\label{fig:bodies}
\end{figure*}

Figure \ref{fig:bodies} shows 3D renderings of the invading structures from scans made after approximately 23 ml of invading fluid was injected, for two different flow rates, 1 and 8 ml/min. As expected, the structures become more dense and occupy a higher saturation for the higher flow rate. The higher the flow rate, the more reasonable the hemisphere assumption becomes.

 \begin{table}[h!]
\caption{Experimental data. For $Q=0.1$ ml/min to $Q=2.0$ ml/min, the corresponding images are taken the moment the invading fluid reaches the bottom of the cell. For $Q=$ 4.0 and 8.0 ml/min, the structures reached the cell walls before reaching the bottom and were imaged at that time. Thus, data for time and injected volume are not strictly comparable. It should also be noted that the times and corresponding volumes for break-through are approximate, as scans where  conducted at two-minute intervals.  $R_c$ measurements are calculated as explained in Fig. \ref{fig:radiusPlot}, with the error as the larger of the difference in $R_c$, from the reference 0.95 saturation threshold to the 0.92 and 0.98 thresholds.}
\centering
\begin{tabular}{|l|r|r|r|}
\toprule
Flow  		        &Percolation    	     	&Injected        &  Measured $R_c$\\
Rate	     		    &Time 			        	&Volume         &  \\
\hline
0.1 ml/min		&73$\pm 2$ min							&7.3$\pm0.2$ ml		&2.1 $\pm$ 0.4 mm\\
0.2 ml/min		&44$\pm 2$ min							&8.8$\pm0.4$ ml		&2.3 $\pm$ 0.2 mm\\
0.3 ml/min		&25$\pm 2$ min							&7.5$\pm0.6$ ml		&4.3 $\pm$ 0.7 mm\\
0.4 ml/min		&32$\pm 2$ min							&12.8$\pm0.8$ ml		&4.6 $\pm$ 2.4 mm\\
0.5 ml/min		&30$\pm 2$ min							&15.0$\pm1.0$ ml		&8.9 $\pm$ 4.5 mm\\
0.6 ml/min		&28$\pm 2$ min							&16.8$\pm1.2$ ml		&12.6 $\pm$ 2.5 mm\\
0.7 ml/min		&32$\pm 2$ min							&22.4$\pm1.4$ ml		&13.5 $\pm$ 2.3 mm\\
0.8 ml/min		&31$\pm 2$ min							&24.8$\pm1.6$ ml		& 11.0 $\pm$ 1.9 mm\\
0.9 ml/min		&25$\pm 2$ min							&22.5$\pm1.8$ ml		& 10.0 $\pm$ 3.3 mm\\
1.0 ml/min		&25$\pm 2$ min							&25.0$\pm2.0$ ml		& 21.0 $\pm$ 2.9 mm\\
1.5 ml/min		&30$\pm 2$ min							&45.0$\pm3.0$ ml		& 18.3 $\pm$ 3.7 mm\\
2.0 ml/min		&17$\pm 2$ min							&34.8$\pm4.0$ ml		& 20.3 $\pm$ 4.1 mm\\
4.0 ml/min		&12$\pm 2$ min							&46.9$\pm8.0$ ml		& 22.1 $\pm$ 3.9 mm\\
8.0 ml/min		&	7$\pm 2$ min							&56.2$\pm16.0$ ml		& 26.6 $\pm$ 3.0 mm\\
\hline
\end{tabular}
\label{table:slow}
\end{table}

Table \ref{table:slow} shows data obtained from images of the experiments. As can be seen, the three lowest flow rates lead to very similar global saturation, while beyond a flow rate of 0.4 ml/min we see a more systematic trend toward the invasion being more stabilized, resulting in higher saturation.

\begin{figure}[h!]
	\centering
	\includegraphics[width=1\linewidth]{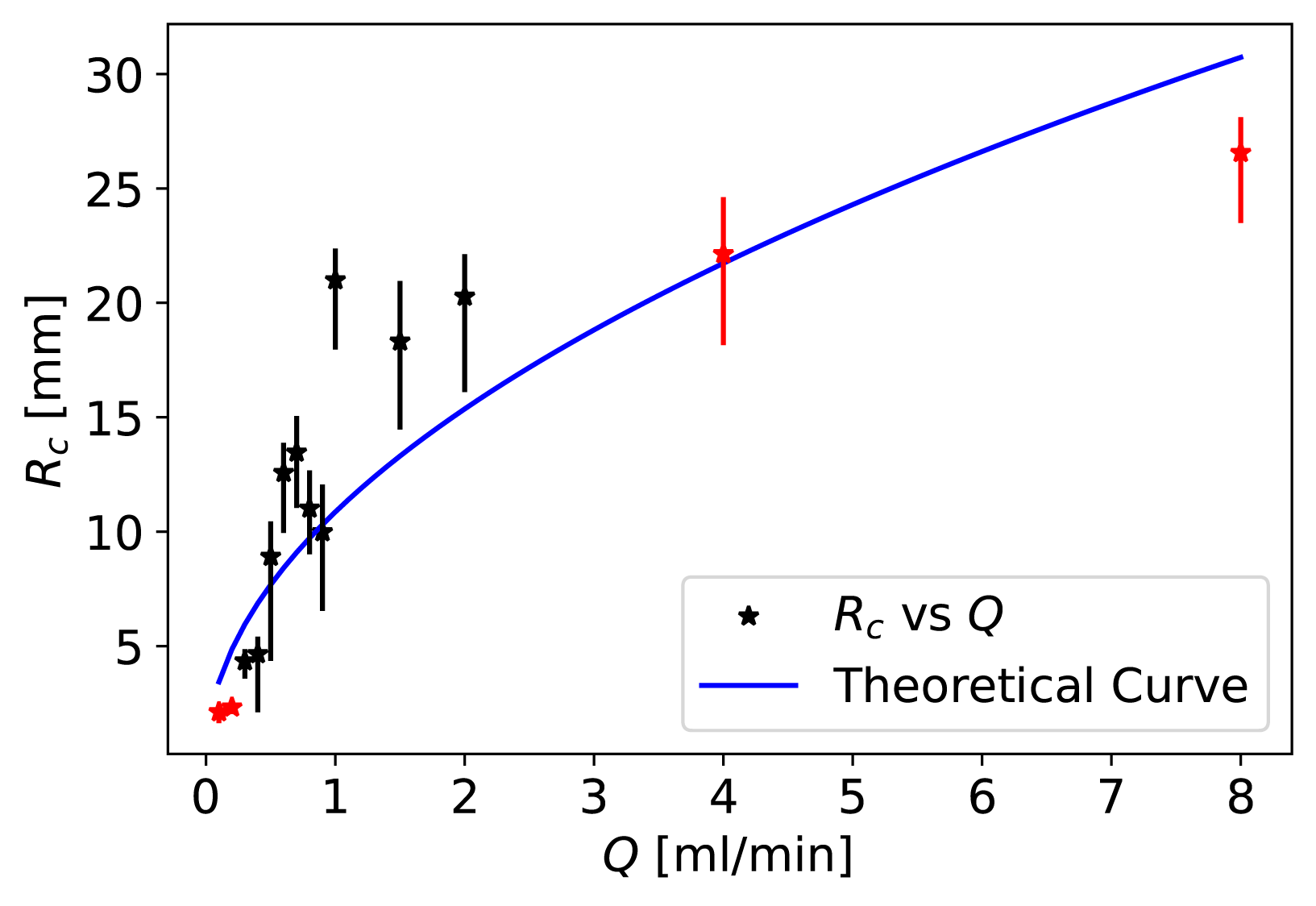}
	\caption{The measured transition radii for the 14 flow rates, and the theoretical prediction, Eq (\ref{eq:Rc}). The red dots are measurements where the the radii are measured below one pore radius and the two fastest experiments, where the cell was not large enough to determine if the transition had occurred. }
	\label{fig:RvsQplot}
\end{figure}

Figure \ref{fig:RvsQplot} shows a plot of  the measured transition radii vs the of flow-rate, together with the theoretical prediction, Eq (\ref{eq:Rc}). 

\section{Discussion and conclusion}\label{sec:disc}
The measured critical radii, $R_c$, fit well with the described theory, but there are not enough experiments, and the considered system is not large enough to conclude firmly and generate statistics over several decades. However, our results are significant in charting the way for experimental work on  flow in 3D porus media. These experiments are inherently difficult, and there will always be limitations. 

In this paper we have described the steps for the construction of a table-top set-up for imaging multiphasic flows in 3D porous media. This set-up is considerably less expensive than traditional methods, such as X-ray or magnetic resonance set-ups and allows for direct optical visualization of flows without any need for \textit{image inversion} algorithms. We have employed the set-up to show how viscous forces can stabilize an invasion front that would be rendered unstable due to an unfavorable density contrast (heavier fluid injected on top of a lighter one). Our set-up allowed for the complete 3D characterization of the invading structure. By considering the balance between the (stabilizing) viscous and (destabilizing) gravitational forces, we have derived an expression for flow-rate dependence of the critical radius, beyond which the front becomes unstable. Although the experimental dataset does not extend far enough for us to accurately measure the dependence of this critical radius on the flow rate, our experiments are in reasonable agreement with the theoretical prediction.

The experimental set-up illustrated here can in principle be further developed to study flows of more fluid phases, the limitation for this being the constraint of matching the index of refraction of the additional fluid phases to the ones already present. As mentioned previously, the limitations due to the imperfect matching of the refraction index can lead to a reduction of image quality, and one must take into account the desired pore-scale resolution when deciding how accurate the index matching needs to be.

Keeping the optical path constant is a critical component of the set-up we describe. This is achieved by tuning the speeds of the laser sheet and the cameras accordingly, as illustrated in Eq. (\ref{eq:vel}). In spite of being a relatively simple step to execute, our experience with the set-up showed early on that this key element has a dramatic effect on the overall quality of the 2D images (and consequently on the quality of the 3D reconstruction).

Our optical scanner adds to the tools of imaging capabilities available to the scientific community. In spite of its relative simplicity and low cost, it does not invalidate other techniques traditionally employed. Each imaging technique has its own applicability range and associated limitations or drawbacks. Here, the advantage of not needing radioactive materials or large facilities is counteracted by the limitation of needing fluids with matching refraction indices. In applications in which this requirement can be met, we believe this technique can be useful for a wide range of systems.

\section*{acknowledgments}
We gratefully acknowledge the support from the University of Oslo and the Research Council of Norway through its Centre of Excellence funding scheme with project number 262644.

\bibliographystyle{apsrev4-1}
 \bibliography{biblio}

\end{document}